\documentclass[twocolumn,showpacs,showkeys,preprintnumbers,amsmath,amssymb,superscriptaddress]{revtex4-1}
\usepackage{graphicx}
\usepackage{natbib}
\usepackage{color}
\usepackage{epstopdf}
  \hypersetup{
   colorlinks,
   linkcolor={red},
   citecolor={blue},
   urlcolor={blue}
  }
 \hypersetup{breaklinks=true}

\begin{document}

\title{Scission configuration of $^{239}$U from yields and kinetic information of fission fragments}
\author{D.~Ramos}
\email[Email address: ]{diego.ramos@ganil.fr}
\affiliation{GANIL, CEA/DRF-CNRS/IN2P3, BP 55027, F-14076 Caen Cedex 5, France}
\affiliation{IPN Orsay, Universit\'e de Paris-Saclay, CNRS/IN2P3, F-91406 Orsay Cedex, France}
\author{M.~Caama\~no}
\affiliation{IGFAE, Universidade de Santiago de Compostela, E-15706 Santiago de Compostela, Spain}
\author{A.~Lemasson}
\affiliation{GANIL, CEA/DRF-CNRS/IN2P3, BP 55027, F-14076 Caen Cedex 5, France}
\author{M.~Rejmund}
\affiliation{GANIL, CEA/DRF-CNRS/IN2P3, BP 55027, F-14076 Caen Cedex 5, France}
\author{H.~Alvarez-Pol}
\affiliation{IGFAE, Universidade de Santiago de Compostela, E-15706 Santiago de Compostela, Spain}
\author{L.~Audouin}
\affiliation{IPN Orsay, Universit\'e de Paris-Saclay, CNRS/IN2P3, F-91406 Orsay Cedex, France}
\author{J.D.~Frankland}
\affiliation{GANIL, CEA/DRF-CNRS/IN2P3, BP 55027, F-14076 Caen Cedex 5, France}
\author{B.~Fern\'andez-Dom\'inguez}
\affiliation{IGFAE, Universidade de Santiago de Compostela, E-15706 Santiago de Compostela, Spain}
\author{E.~Galiana-Bald\'o}
\affiliation{IGFAE, Universidade de Santiago de Compostela, E-15706 Santiago de Compostela, Spain}
\affiliation{LIP Lisboa, 1649-003 Lisbon, Portugal}
\author{J.~Piot}
\affiliation{GANIL, CEA/DRF-CNRS/IN2P3, BP 55027, F-14076 Caen Cedex 5, France}
\author{C.~Schmitt}
\affiliation{IPHC Strasbourg, Universit\'e de Strasbourg-CNRS/IN2P3, F-67037 Strasbourg Cedex 2, France}
\author{D.~Ackermann}
\affiliation{GANIL, CEA/DRF-CNRS/IN2P3, BP 55027, F-14076 Caen Cedex 5, France}
\author{S.~Biswas}
\affiliation{GANIL, CEA/DRF-CNRS/IN2P3, BP 55027, F-14076 Caen Cedex 5, France}
\author{E.~Clement}
\affiliation{GANIL, CEA/DRF-CNRS/IN2P3, BP 55027, F-14076 Caen Cedex 5, France}
\author{D.~Durand}
\affiliation{LPC Caen, Universit\'e de Caen Basse-Normandie-ENSICAEN-CNRS/IN2P3, F-14050 Caen Cedex, France}
\author{F.~Farget}
\affiliation{LPC Caen, Universit\'e de Caen Basse-Normandie-ENSICAEN-CNRS/IN2P3, F-14050 Caen Cedex, France}
\author{M.O.~Fregeau}
\affiliation{GANIL, CEA/DRF-CNRS/IN2P3, BP 55027, F-14076 Caen Cedex 5, France}
\author{D.~Galaviz}
\affiliation{LIP Lisboa, 1649-003 Lisbon, Portugal}
\author{A.~Heinz}
\affiliation{Chalmers University of Technology, SE-41296 G\"oteborg, Sweden}
\author{A.~Henriques}
\affiliation{CENBG, IN2P3/CNRS-Universit\'e de Bordeaux, F-33175 Gradignan Cedex, France}
\author{B.~Jacquot}
\affiliation{GANIL, CEA/DRF-CNRS/IN2P3, BP 55027, F-14076 Caen Cedex 5, France}
\author{B.~Jurado}
\affiliation{CENBG, IN2P3/CNRS-Universit\'e de Bordeaux, F-33175 Gradignan Cedex, France}
\author{Y.H.~Kim}
\thanks{Present address: ILL, F-38042 Grenoble Cedex 9, France}
\affiliation{GANIL, CEA/DRF-CNRS/IN2P3, BP 55027, F-14076 Caen Cedex 5, France}
\author{P.~Morfouace}
\thanks{Present address: CEA, DAM, DIF, F-91297 Arpajon, France}
\affiliation{GANIL, CEA/DRF-CNRS/IN2P3, BP 55027, F-14076 Caen Cedex 5, France}
\author{D.~Ralet}
\affiliation{CSNSM, CNRS/IN2P3, Universit\'e de Paris-Saclay,F-91405 Orsay, France}
\author{T.~Roger}
\affiliation{GANIL, CEA/DRF-CNRS/IN2P3, BP 55027, F-14076 Caen Cedex 5, France}
\author{P.~Teubig}
\affiliation{LIP Lisboa, 1649-003 Lisbon, Portugal}
\author{I.~Tsekhanovich}
\affiliation{CENBG, IN2P3/CNRS-Universit\'e de Bordeaux, F-33175 Gradignan Cedex, France}
\date{\today}

\begin{abstract}

The simultaneous measurement of the isotopic fission-fragment yields and fission-fragment velocities of $^{239}$U has been performed for the first time. The $^{239}$U fissioning system was produced in one-neutron transfer reactions between a $^{238}$U beam at 5.88 MeV/nucleon and a $^{9}$Be target. The combination of inverse kinematics at low energy and the use of the VAMOS++ spectrometer at the GANIL facility allows the isotopic identification of the full fission-fragment distribution and their velocity in the reference frame of the fissioning system. The proton and neutron content of the fragments at scission, their total kinetic and total excitation energy, as well as the neutron multiplicity were determined. Information from the scission point configuration is obtained from these observables and the correlation between them. The role of the octupole-deformed proton and neutron shells in the fission-fragment production is discussed. 

\end{abstract}

\keywords{}

\pacs {}

\maketitle

\section{INTRODUCTION}

The fission reaction is one of the most complex and, consequently, one of the less understood mechanisms in nuclear physics~\cite{Schmidt2018}. Its complexity lies in the fact that several properties of nuclear matter and nuclear dynamics play an important role at different stages of the process, and the resulting products cannot be fully explained without taking into account all these ingredients and the continuous interplay between them~\cite{Schunck2016}. For instance, at low excitation energy, the internal nuclear structure defines the more relevant properties of the final fission fragment distributions~\cite{wil76,Scamps2018} while the dissipative dynamics of the process is the responsable for the enhancement of intrinsic excitations leading to the stochastic rupture of nucleon pairs along the path from the saddle to the scission point~\cite{Steinhauser1998,Rejmund2000,Caamano2011}. 

The many-body problem introduces additional constrains in the modeling of the process: the microscopic treatment of the problem, taking into account the full coupling between intrinsic and collective degrees of freedom, is at the limit of the current supercomputers capabilities for one single fission event~\cite{Simenel14,Bulgac2016,Scamps2018,Bulgac19}. Approximations are needed to disentangle both intrinsic and collective contributions in order to reproduce fission-fragment distributions~\cite{Regnier16,Zhao19}. Additional theoretical approaches are based on the phenomenological description of the problem using a multidimensional parametrization of the system~\cite{Moller2001,Ward2017} or the statistical study at the scission point~\cite{Lemaitre19}. In this framework, model predictions need to be compared with a set of experimental observables wide enough to probe the correct approach of the problem.

The proton and neutron content of the fission fragments at scission are suitable observables for exploring the impact of the intrinsic structure; the kinetic energy of the fission fragments gives relevant information about the shape of the system at scission; and the neutron multiplicity is a good indicator of the temperature of the system and the fragments because the neutron evaporation is the main release mechanism of excitation energy.  

\begin{figure*}[!]
\includegraphics[width=0.49\textwidth]{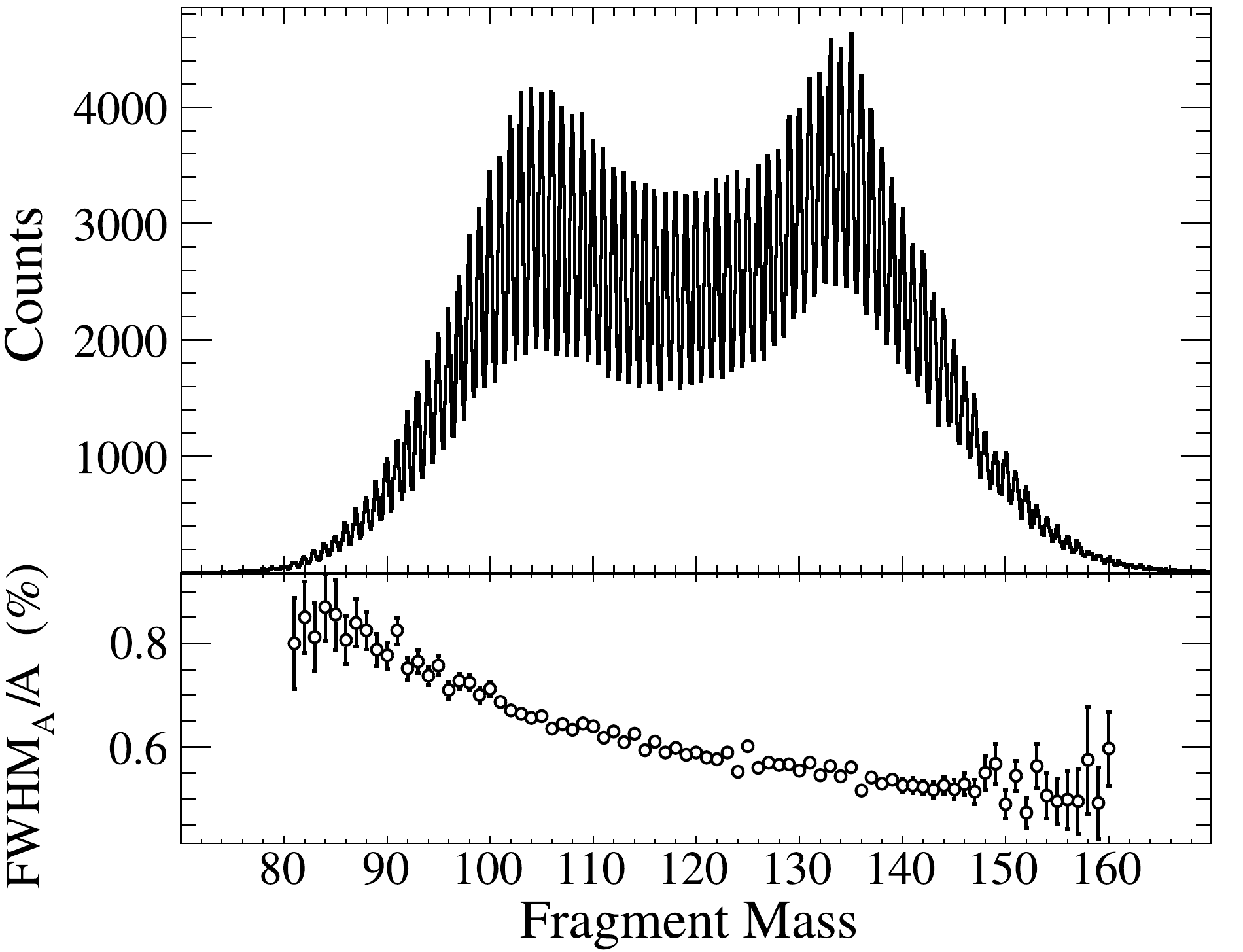}
\includegraphics[width=0.49\textwidth]{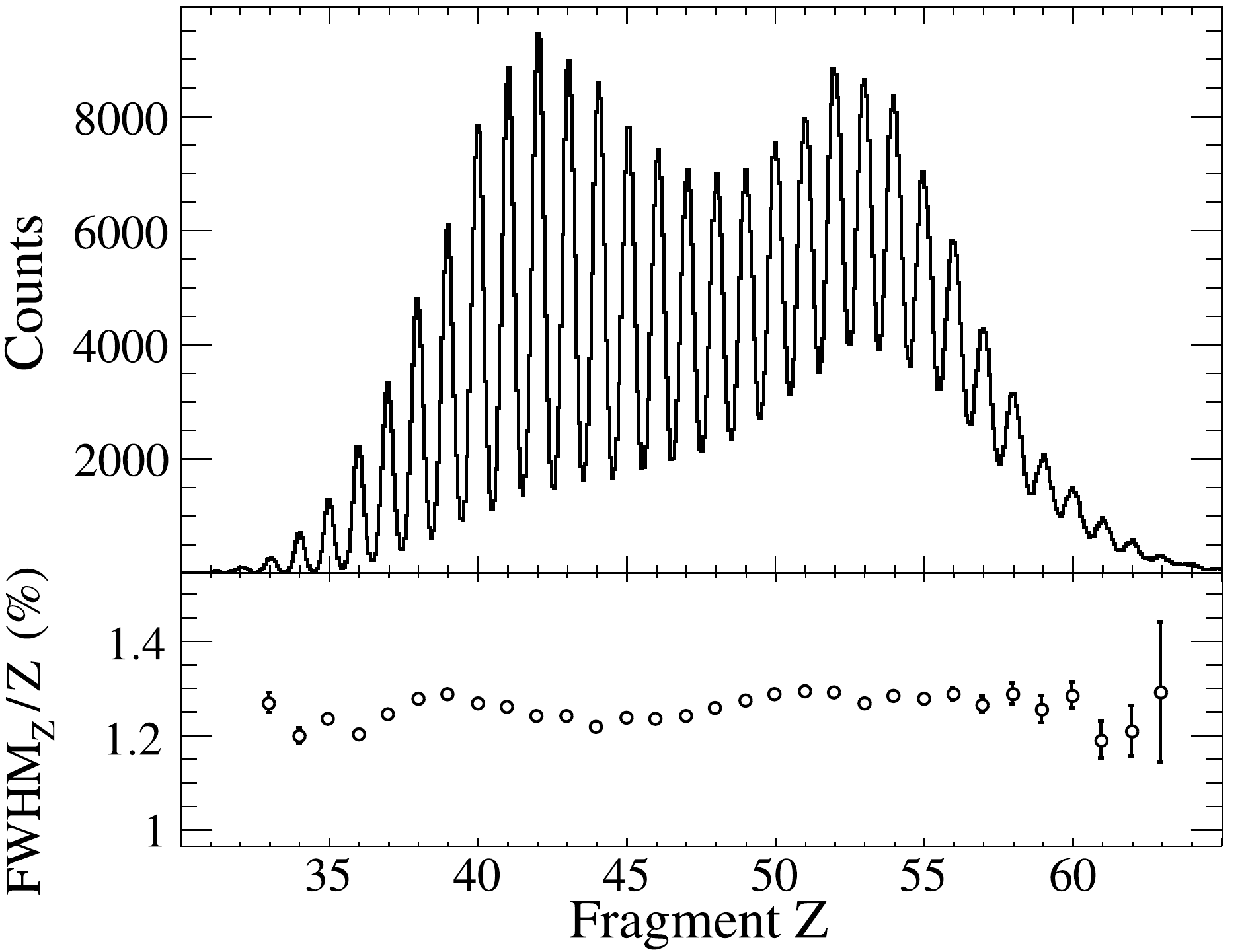}
\caption{Fission-fragment mass and atomic number distributions measured in the VAMOS++ focal-plane detection-setup. Lower panels show the resolution of the measurements in relative full width at half maximum.}
\label{fig::FFID}
\end{figure*}

The experimental access to all these observables is still limited. In particular, the neutron content of the fission fragments at scission is challenging because the fragments would need to be isotopically identified before neutron evaporation. Nevertheless, experimental developments have been performed during the last years in order to extend the number of available observables and systems. Among others, surrogate reactions give the opportunity to explore fission in short-lived radioactive isotopes~\cite{Escher12,Kessedjian2010,Carme2014,Nishio2015,Leguillon16,Ducasse16,Hirose2017}, while the inverse kinematics technique allows the measurement of the proton content of the full fragment distribution~\cite{Steinhauser1998,Schmidt2000,Bockstiegel2008}. Furthermore, the use of inverse kinematics coupled with magnetic spectrometers allows the simultaneous measurement of both the mass and proton content of the full fission-fragment distributions~\cite{Caamano2013a,JLRS2015,Pellereau2017,Ramos2019,chatillon19}.

The present paper reports on the latest results of the ongoing fission campaign in inverse kinematics with the magnetic spectrometer VAMOS++ at GANIL. The isotopic fission yields --- subject of a previous publication~\cite{RamosPRL2019} --- as well as the velocity of the fission fragments of $^{239}$U at 8.3 MeV of excitation energy were measured. The average neutron content of the fission fragments at scission, obtained from the fragment velocity measurements, the neutron multiplicity, and the total kinetic energy distributions are presented. These results allow us to explore the impact of the octupole-deformed shells~\cite{Scamps2018} and the scission configuration of the system. 

The paper is organized as follows: Section~\ref{sec_exp} summarizes the experimental setup used for this measurement; Section~\ref{sec_obs} presents the fission-fragment observables relevant for the results discussed in this paper; Section~\ref{sec_res} shows the results obtained regarding $^{239}$U; Section~\ref{sec_dis} presents a discussion based on the comparison of the present data with a previous measurement on $^{240}$Pu; finally Section~\ref{sec_con} presents the conclusions of this work.    

\section{EXPERIMENTAL SETUP}
\label{sec_exp}
The experiment was performed at the GANIL facility where a $^{238}$U beam at 5.88~MeV/nucleon, with an intensity of $10^9$ pps, was used. The beam impinged on a 500-$\mu$g/cm$^{2}$ thick $^{9}$Be target, populating $^{239}$U by one-neutron transfer reactions [$^{9}$Be( $^{238}$U, $^{239}$U) $^{8}$Be]. Once the transfer reaction takes place, if the excitation energy of $^{239}$U is high enough to overcome the fission barrier ($B_{f} = 6.4$~MeV~\cite{Bj1980}), the system undergoes fission and both fragments are emitted at forward angles within a cone of $\sim$30$^\circ$ with respect to the beam and with energies ranging from 2 to 10~MeV/nucleon. 

The target-like recoil $^{8}$Be is most likely emitted at $\sim$43$^\circ$ and decays into two $\alpha$-particles. The fissioning $^{239}$U nuclei are identified by detecting the $\alpha$-$\alpha$ coincidence in SPIDER~\cite{Carme2014}, a double-sided segmented silicon telescope placed 31~mm downstream from the target. The distribution of excitation energy of the fissioning system is obtained, event-by-event, from the reconstruction of the binary reaction as discussed in Ref.~\cite{RamosPRL2019}. The mean excitation energy results in 8.3~MeV with a standard deviation of 2.7~MeV.

For each fission event, one of the fission fragments passes through the VAMOS++ spectrometer~\cite{Rejmund2011} and is fully identified at its focal plane setup. The atomic number of the fission fragment is obtained with the $\Delta E$-$E$ measurement from a segmented ionization chamber, filled with CF$_4$ gas at 100 mbar. The mass number is obtained from the magnetic rigidity, reconstructed from the trajectory followed by the fragment inside the spectrometer. In addition, the velocity of the fragment is deduced from the time-of-flight between two multiwire chambers placed before and after the VAMOS++ spectrometer, respectively, and the reconstructed path followed by the fragment between the target and the focal plane position of the spectrometer, with a nominal value of 760~cm. The emission angles of the fragment are also measured in the DP-MWPC placed at 17~cm downstream from the target~\cite{Vandebrouck2016}. The phase space of the measured fission fragments covers between 5 and 10 MeV/nucleon in energy and from 7 to 28$^\circ$ in polar angle. This range is achieved by using VAMOS++ rotated at 21.5$^\circ$ and 14$^\circ$ with respect to the beam axis, and with central magnetic rigidities of 1.1 and 1.24~Tm, respectively.  

With this setup configuration, the mass and atomic numbers of the full fission-fragment distribution are unambiguously assigned, and their velocity vector is measured. Further details on  VAMOS++, along with typical performances for the fission-fragments detection are given in Ref.~\cite{Kim2017}.

The fission-fragment phase-space covered by the VAMOS++ acceptance, and the stopped low-energy fragments, reduce the polar-angle coverage, which evolves with the fragment mass from 15\% to 40\%. The azimuthal-angle coverage depends on both, the magnetic rigidity and the polar angle, and ranges from 5\% to 20\%. 

\section{FISSION-FRAGMENT OBSERVABLES}
\label{sec_obs}
Upper panels of Fig.~\ref{fig::FFID} present the mass and atomic number distributions of the fission fragments, all reaction channels included, measured at the focal plane of VAMOS++. Masses ranging from 80 up to 160 and atomic numbers from 32 up to 64 are identified. The feeding of intermediate and very heavy elements is produced by fusion-fission reactions between the $^{238}$U beam and the $^{9}$Be target, which are ten times more likely than any transfer channel. This contribution is subtracted with the selection of the fissioning system in SPIDER and the analysis procedure, as discussed in Ref.~\cite{RamosPRL2019}. The resolution achieved in mass and atomic numbers are presented in the lower panels of Fig.~\ref{fig::FFID}, in terms of the relative full width at half maximum as a function of the mass and atomic numbers. The mass resolution is limited by the time-of-flight resolution, and it ranges from 0.8\%, for lighter fragmentes, down to 0.5\%, for heavier ones. The atomic number resolution has an average value lower than 1.3\%, limited by the intrinsic resolution of the ionization chamber. 

The measured velocity is corrected by the slow down of the fragment in the target, according to the stopping power of each ion as a function of its nuclear charge, mass number, and measured velocity, as discussed in~\cite{DiegoPhD}. 

\begin{figure}[h!]
\includegraphics[width=0.49\textwidth]{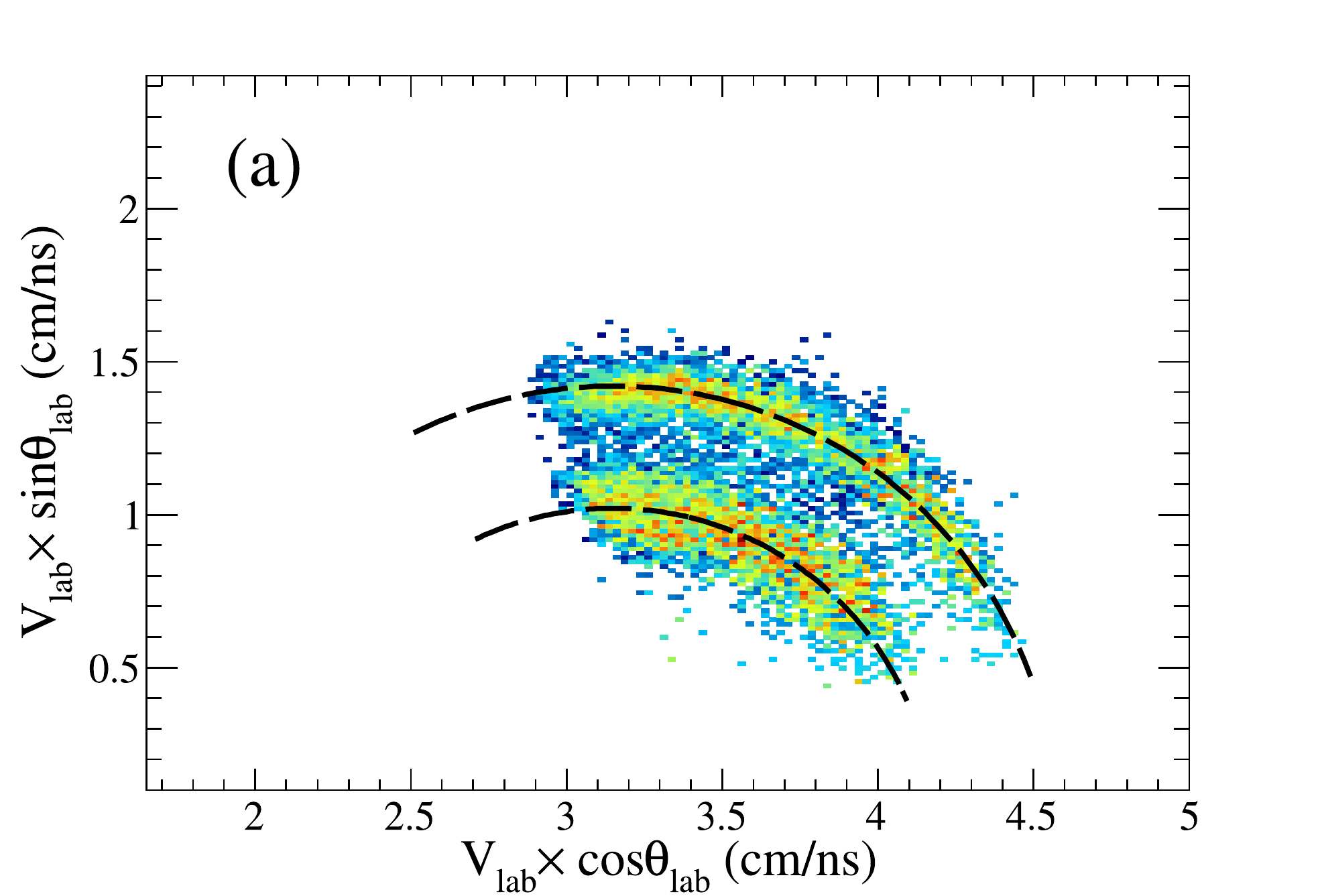}
\includegraphics[width=0.49\textwidth]{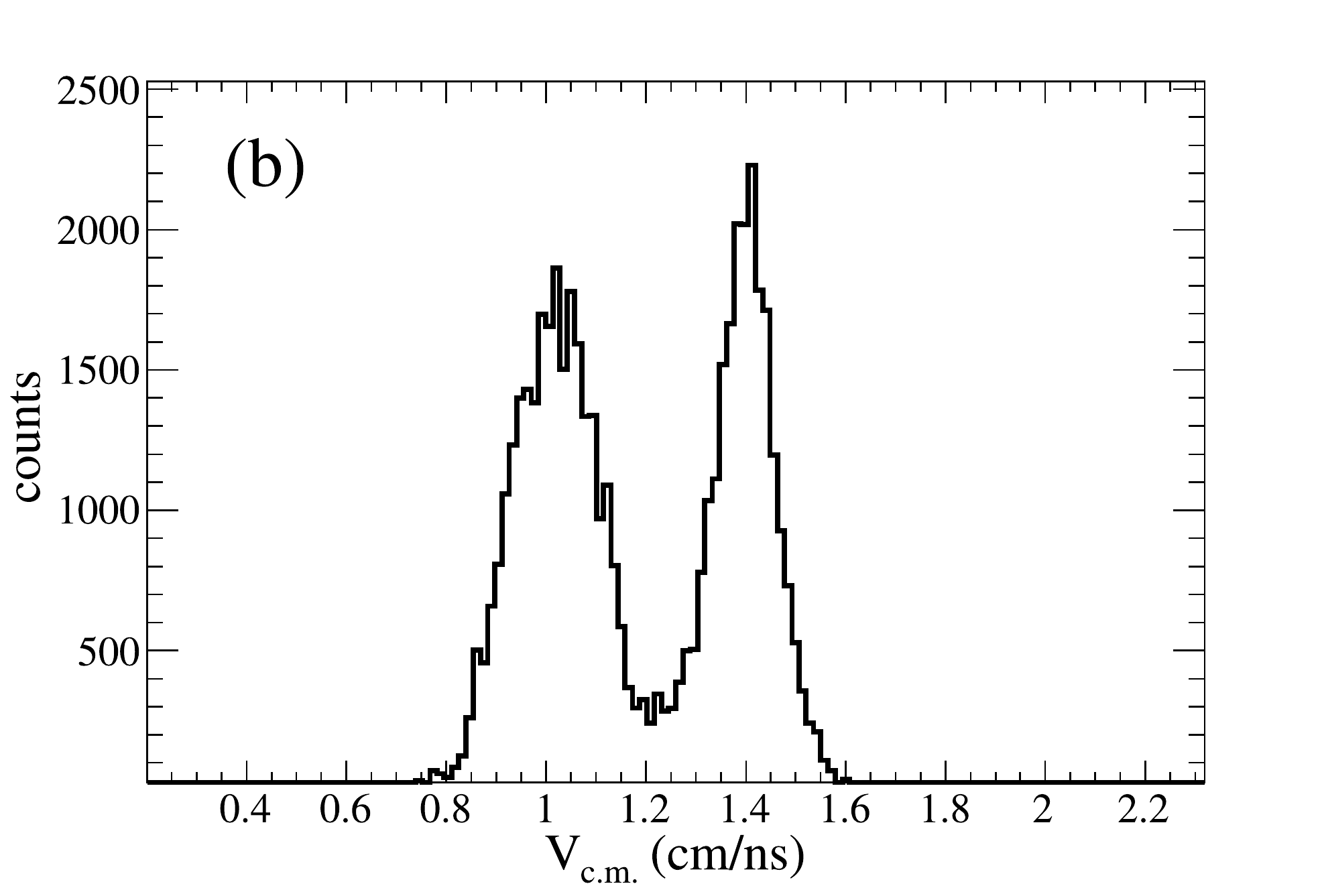}
\caption{(Color online) Fission fragment velocity of $^{239}$U: (a) Velocity of fission fragments in the laboratory reference frame. The horizontal/vertical axis corresponds to the projection of the velocity parallel/perpendicular to the forward trajectory. The dashed lines indicate the mean value of the heavy and light fragment. (b) Distribution of the fission-fragment velocity reconstructed in the reference frame of the fissioning system.}
\label{fig::Vff}
\end{figure}

Figure~\ref{fig::Vff} (a) shows the velocity distribution of the fission fragments of $^{239}$U in the laboratory reference frame (\textit{lab}). The vertical and horizontal axes represent the parallel and perpendicular components of the velocity vector with respect to the direction of the fissioning system. Two well-separated distributions are identified as corresponding to the light and heavy fragment groups. Dashed lines represent the mean value of each distribution. Each line describes a circumference whose radius corresponds to the velocity of the fragments in the reference frame where the fissioning system is at rest (\textit{c.m.}). 
The velocity of the fragments in the \textit{c.m.} reference frame ($v_{c.m.}$), calculated by using the Lorentz formalism,is presented in Fig.~\ref{fig::Vff} (b). The wider distribution centered around 1~cm/ns corresponds to the heavy fragment group, while the narrower distribution centered around 1.4~cm/ns corresponds to the light fragment group.  

The uncertainty in $v_{c.m.}$ is determined from the resolution of the time-of-flight measurement, between 400 and 600~ps, the resolution of the measurement of the fragment angles, of 0.14$^{\circ}$, and the uncertainty in the reconstruction of the path traveled by the fragment, estimated to 0.1\%. The total uncertainty of the reconstructed $v_{c.m.}$ is less than 1.3\%.  

\section{EXPERIMENTAL RESULTS}
\label{sec_res}

\begin{figure}[t!]
\includegraphics[width=0.5\textwidth]{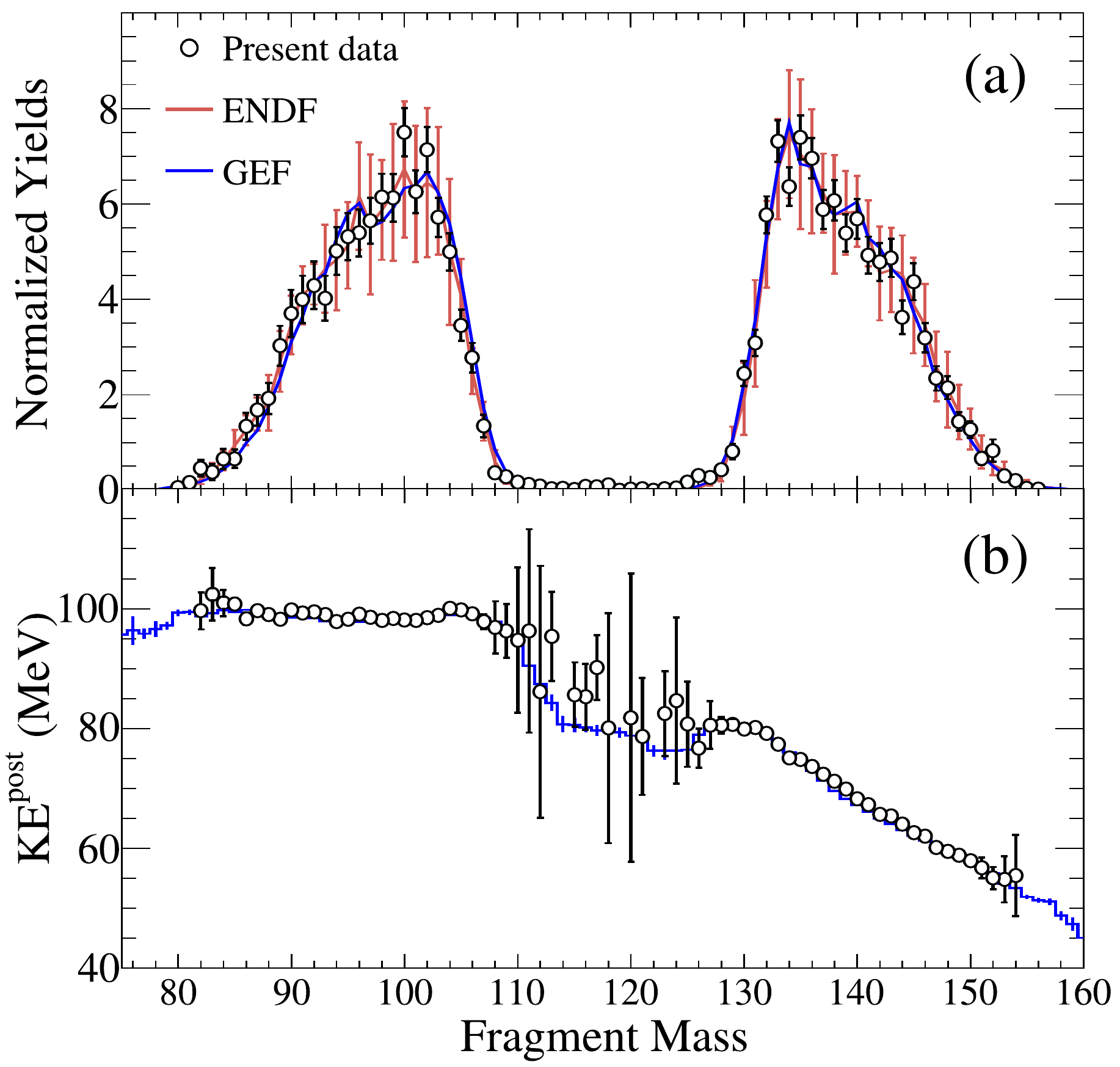}
\caption{(Color online) Yields and kinetic energies of fission fragments of $^{239}$U as a function of post-neutron-evaporation mass: (a) Post-neutron-evaporation mass yields. Present data (circles) are compared with evaluated data from ENDF/B-VIII.0~\cite{ENDF} (red lines) and with the GEF (v. 2019/1.1)~\cite{GEF} (blue line). (b) Post-neutron-evaporation kinetic energy of fragments as a function of post-neutron-evaporation mass. Present data (circles) are compared with the GEF calculation (blue lines).}
\label{fig::YA}
\end{figure}

The post-neutron-evaporation mass yields $Y(A^{post})$ of $^{239}$U are obtained from the isotopic yields~\cite{RamosPRL2019} as the sum on the different elements of every mass. The mass yields are presented in Fig.~\ref{fig::YA} (a), where present data (circles) are compared with 500 keV-neutron-induced fission from the ENDF/B-VIII.0 evaluation~\cite{ENDF} (red lines) and with the GEF code~\cite{GEF} (blue line). Present data are in good agreement with both, evaluation and model, showing a clear asymmetric fission with a very low production at symmetry. Present data may contribute to constrain the evaluation showing systematically smaller uncertainties. The systematic uncertainties, ranging from 2\% in the heavier fragments up to 10\% in the lighter ones, include those from the determination of the spectrometer acceptance, the relative spectrometer settings normalization, and the contamination subtraction from fusion-fission. A remaining contamination from other fissioning systems different from $^{239}$U was estimated to be lower than 0.9\%.

The post-neutron-evaporation kinetic energy of fission fragments ($KE^{post}$) is calculated from $v_{c.m.}$ and the post-neutron-evaporation fragment mass ($M^{post}$), approximated to the mass number ($M^{post}\approx u\times A^{post}$) with an error smaller than 0.1\%. 

For each mass, $KE^{post}$ is determined as a function of $A^{post}$ as the average contribution of the different elements:
\begin{equation}
KE^{post}(A^{post}) = \frac{\sum_{Z}\left[KE^{post}(Z,A^{post})\times Y(Z,A^{post})\right]}{\sum_{Z}Y(Z,A^{post})}.
\end{equation}

Figure ~\ref{fig::YA} (b) shows the post-neutron-evaporation kinetic energy as a function of the fission-fragment mass. Present data (circles) show large fluctuations at symmetry because of the reduced production in this region of the strong asymmetric fission of $^{239}$U at this energy. In the following, this region will be excluded from the discussion because it is not statistically conclusive. Present data, in perfect agreement with the GEF calculation, show a constant kinetic energy for the light group and a decreasing kinetic energy with increasing mass number in the heavy group.

\begin{figure}[t!]
\includegraphics[width=0.5\textwidth]{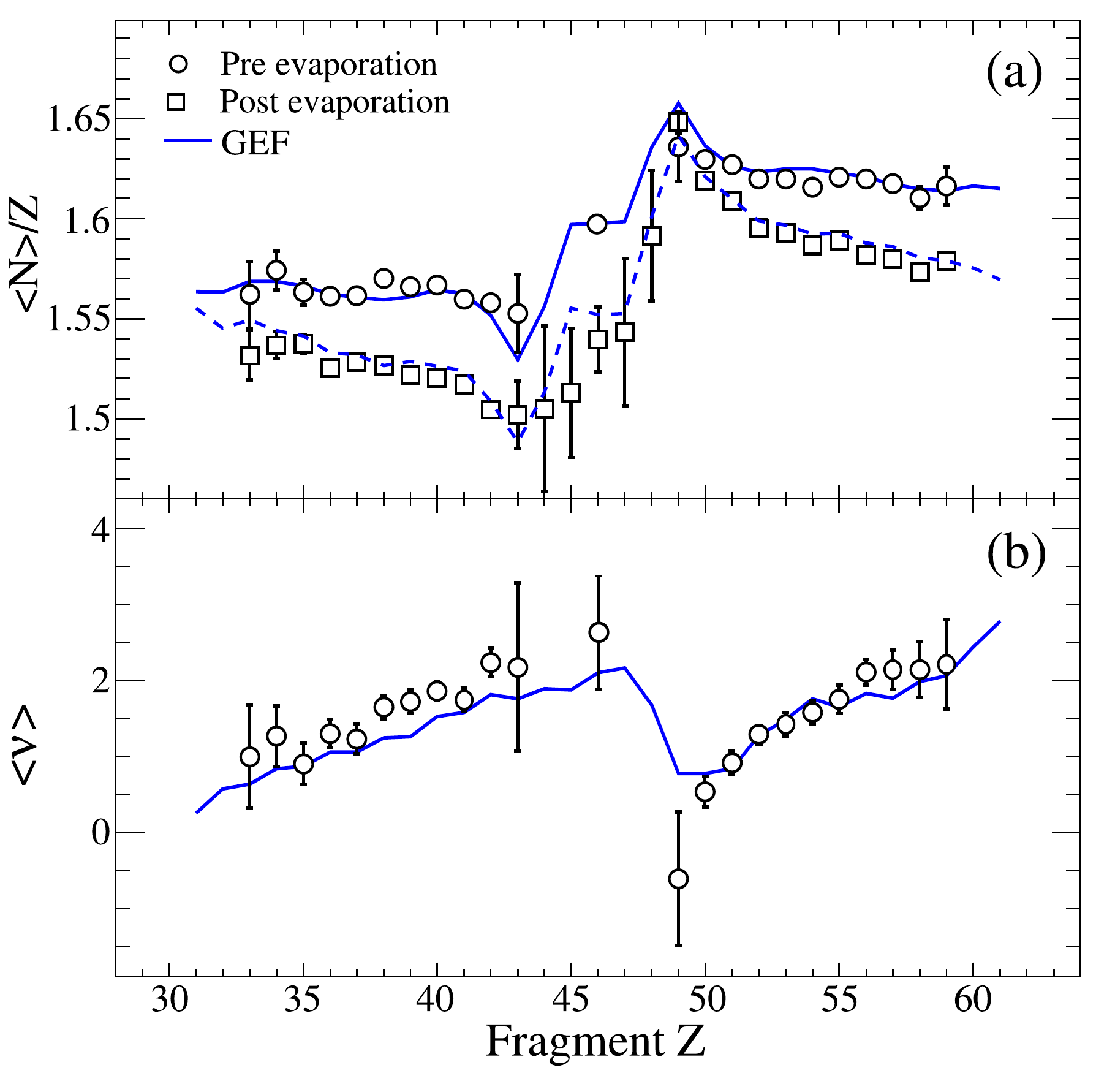}
\caption{(Color online) Neutron excess and neutron multiplicity of fission fragments of $^{239}$U: (a) Average neutron excess of fission fragments pre- and post-neutron-evaporation as a function of the atomic number. (b) Average neutron multiplicity as a function of the atomic number. Present data (symbols) are compared with the GEF (v. 2019/1.1) calculation~\cite{GEF} (lines).}
\label{fig::NZ}
\end{figure}

The access to $v_{c.m.}$ allows to obtain the pre-neutron-evaporation masses of complementary fragments ($M^*_{1,2}$) by applying momentum conservation: 
\begin{equation}
\frac{M^*_1}{M^*_2}=\frac{\gamma_{c.m._2} v_{c.m._2}}{\gamma_{c.m._1} v_{c.m._1}}.
\end{equation}

In the present experiment, only one of the two fragments was fully identified for each fission event, therefore the masses are determined in average as a function of the atomic number of the fragments, as disussed in Ref.~\cite{Caamano15}. Average fission-fragment mass numbers before neutron evaporation $\left(\langle A^*_{1,2} \rangle\right)$ are calculated as 
\begin{equation}
\begin{aligned}
&\langle A^*_{1} \rangle(Z_{1}) = A^{FIS}\frac{\langle \gamma_{c.m._2}v_{c.m._2}\rangle(Z_2)}{\langle\gamma_{c.m._1} v_{c.m._1}\rangle(Z_1)+\langle\gamma_{c.m._2}v_{c.m._2}\rangle(Z_2)}, \\
&\langle A^*_{2}\rangle(Z_{2}) = A^{FIS} - \langle A^*_{1} \rangle(Z_{1}),
\end{aligned}
\end{equation}
where $A^{FIS} = \langle A^*_{1} \rangle + \langle A^*_{2} \rangle$ and $Z^{FIS}=Z_1+Z_2$ are the mass and atomic numbers of the fissioning system, respectively. 

The approximations performed in this calculation, namely the assumption of constant fragment velocity before and after neutron evaporation, the ratio equivalence $M^*_1/M^*_2\approx A^*_1/A^*_2$, and the factorization of mean values $\langle A^*\times\gamma_{c.m.}\times v_{c.m.}\rangle\approx\langle A^*\rangle\times\langle\gamma_{c.m.}\times v_{c.m.}\rangle$, bring a final mass-number error smaller than 0.4\%. 

In our calculations, the mass number of the fissioning system was $A^{FIS}=238.92$, obtained from the GEF code, resulting from an average evaporation of 0.08 neutrons at the present energy.

From the average mass number before neutron evaporation, the neutron excess of the fission fragments at the scission point can be determined as a function of the atomic number: $\langle N^*\rangle/Z = \left(\langle A^*\rangle-Z\right)/Z$. These data are presented in Fig.~\ref{fig::NZ} (a) (circles), together with the neutron excess after neutron evaporation (squares), obtained from the measured isotopic-fission yields~\cite{RamosPRL2019}.

The data at the scission point show an enhancement of the neutron content around $Z\sim50$. This indicates that there is a structural effect around Sn that favors a specific sharing of neutrons leading to a heavy fragment with $N\sim82$. The data are compared with the GEF calculation (blue lines), showing a similar behavior. 

The average neutron multiplicity as a function of the atomic number can be obtained from the difference between both pre- and post-neutron-evaporation neutron excess. These data are shown in Fig.~\ref{fig::NZ} (b) (circles) compared with GEF (blue line). Both describe a sawtooth shape with a minimum around $Z\sim50$ that it is consistent with the minimum observed in previous measurements around $A\sim132$~\cite{Walsh77,Nishio95,Nishio98} for different systems. This can be explained as due to the magicity of Sn that prevents a high excitation energy from large deformation and hence, the neutron evaporation is reduced. In the light fragment region, the GEF calculation systematically underestimates the neutron multiplicity. 

\begin{figure}[t!]
\includegraphics[width=0.5\textwidth]{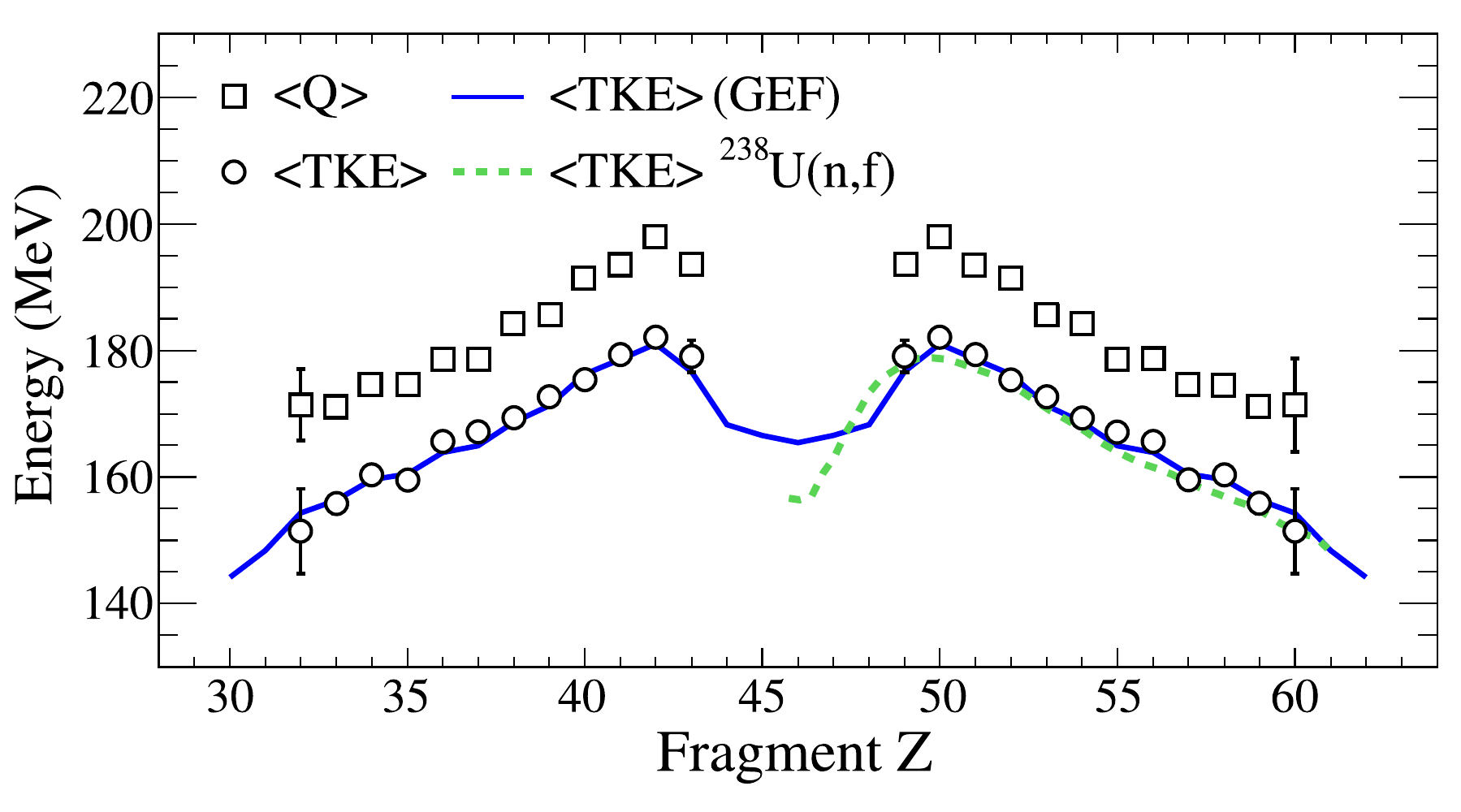}
\caption{(Color online) Average total kinetic energy of fission fragments of$^{239}$U as a function of the fission-fragment atomic number. Present data (circles) are compared with previous measurements from neutron-induced fission~\cite{Vives00} (dotted green line) and with GEF (v. 2019/1.1)~\cite{GEF} (solid blue line). The average $Q$-value of the reaction as a function of the atomic number is also presented (squares).} 
\label{fig::TKE}
\end{figure}

Additional information about the scission point configuration is gathered from the total kinetic energy of the fragments ($TKE$), determined from $v_{c.m.}$ and the pre-neutron-evaporation masses. In this case, the $TKE$ is also determined as an average as a function of the atomic number of the fragments, 
 \begin{equation}
 \begin{aligned}
 &\langle TKE\rangle(Z_1) = \langle TKE\rangle(Z_2)= \\
 &\langle M^*_1\rangle\left[\langle\gamma_{c.m._1}\rangle-1\right](Z_1)+ \langle M^*_2\rangle\left[\langle\gamma_{c.m._2}\rangle-1\right](Z_2),
 \end{aligned}
 \end{equation}
where $\langle M^*_{1,2}\rangle$ are the average ground-state masses of the fragments at the scission point, determined as described in Ref.~\cite{Caamano15}.

Figure~\ref{fig::TKE} presents the $\langle TKE\rangle$ measurements of this work (circles) compared with GEF (solid blue line) and with previous measurements from 1.8 MeV-n-induced fission~\cite{Vives00} (dotted green line). The $\langle TKE\rangle$ of this previous measurement was obtained as a function of the mass of the fragments and, in order to compare both sets of data, the mass number is translated into $Z$ using the $N/Z$ of GEF. A good agreement between present data and both GEF and the previous measurement is observed. They reproduce the same shape with a maximum at $Z=50$, which suggests a compact configuration at scission. Contrary, in a previous work on neutron-deficient actinides~\cite{Bockstiegel2008}, the maximum of $\langle TKE\rangle$ was observed at $Z=52$. This difference can be explained by the small contribution of the symmetric fission component in the present data that prevents a reduction of the $\langle TKE\rangle$ around $Z=50$, thus the characteristic high $TKE$ of the compact configuration is revealed for the present system.

The $\langle Q\rangle$-value $=M^{FIS}-\langle M^*_1\rangle-\langle M^*_2\rangle$ of the reaction is also shown in Fig.~\ref{fig::TKE} (squares) for completeness. It exhibits a similar behavior as the $\langle TKE\rangle$ but with values around 14~MeV higher. The total excitation energy of the fragments ($TXE$), that results from the difference between both, $Q$-value and $TKE$, is discussed in the next section. 

\begin{figure}[t!]
\includegraphics[width=0.5\textwidth]{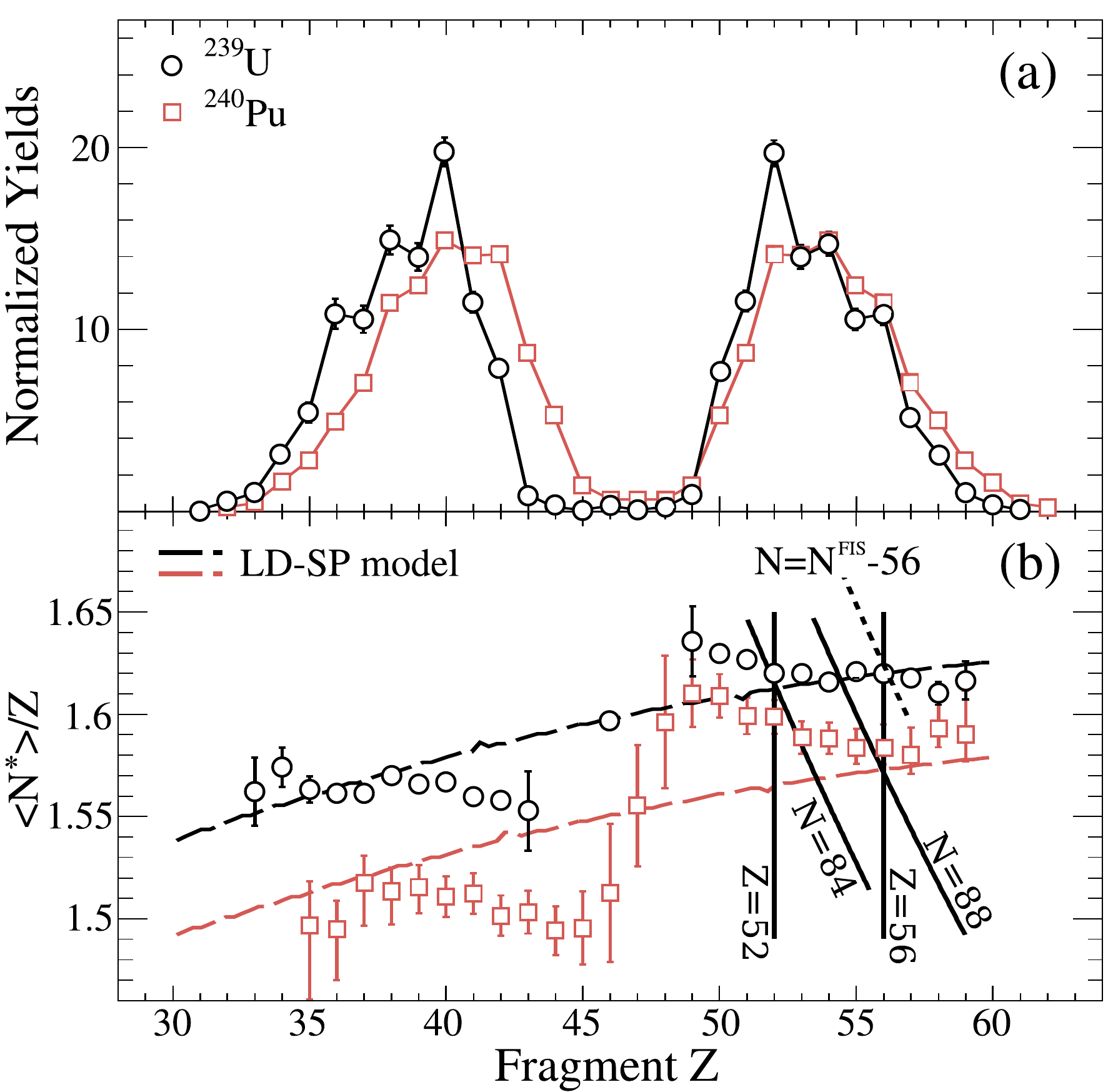}
\caption{(Color online) Elemental fission yields and neutron excess of $^{239}$U and $^{240}$Pu: (a) Normalized yield as a function of the atomic number. Present data (black circles) are compared with a previous measurement of $^{240}$Pu~\cite{Ramos2018} (red squares). (b) Pre-neutron-evaporation neutron excess of fission fragments as a function of the atomic number. Present data (black circles) are compared with a previous measurement of $^{240}$Pu~\cite{Caamano15} (red squares) and with a scission-point model based on the liquid-drop behaviour (dashed lines, colors indicate the systems). Octupole-deformed shells $Z=52,56$ and $N=56,84,88$ are indicated by solid black lines (see text for details).}
\label{fig::NZ_COMP}
\end{figure}

\section{DISCUSSION}
\label{sec_dis}

In this section, general trends of the fission process are discussed with the comparison between the present data of $^{239}$U and a previous measurement of $^{240}$Pu at 9~MeV of excitation energy~\cite{Caamano15}.

Figure~\ref{fig::NZ_COMP} presents the fission yields (a) and the neutron excess of fission fragments at the scission point (b), as a function of the fragment proton number for $^{239}$U (black circles) and $^{240}$Pu (red squares). A scission-point model description of the neutron excess based on a liquid-drop behavior is also shown for both systems (dashed lines).  

In both cases, the fragment production cannot be explained by means of the spherical closed shells $Z=50$ and $N=82$ because, despite of being the region where the $\langle N^{*}\rangle /Z$ differs the most from the liquid-drop trend, the production of $Z=50$ remains relatively low compared with $Z\in [52,56]$.

The proton numbers $Z=52,56$ were recently reported as octupole-deformed shells, together with neutron numbers $N=56,84,88$~\cite{Scamps2018, Scamps2019}. The effect of these octupole-deformed shells appears in the data as an enhanced production of some elements due to the combined effect of proton and neutron shells. The neutron excess (Fig.~\ref{fig::NZ_COMP}(b)) shows that fragments with $Z=52$ are produced with $N=84$ in $^{239}$U, boosting the production around $^{136}$Te; while in $^{240}$Pu, the $N=84$ shell is shared between $Z=52$ and $Z=53$, which explains the yield difference between both systems. The $N=88$ shell is produced with $Z=56$ for $^{240}$Pu while $N=56$ in the light fragment appears with $Z=56$ in the heavy one for $^{239}$U. This also can explain the production around $^{144}$Ba in $^{240}$Pu and around $^{147}$Ba in $^{239}$U. The yields between $Z=52$ and  $Z=56$ appear by the overlap of the natural width of the shells, due to the stochastic nature of the process.

\begin{figure}[t!]
\includegraphics[width=0.5\textwidth]{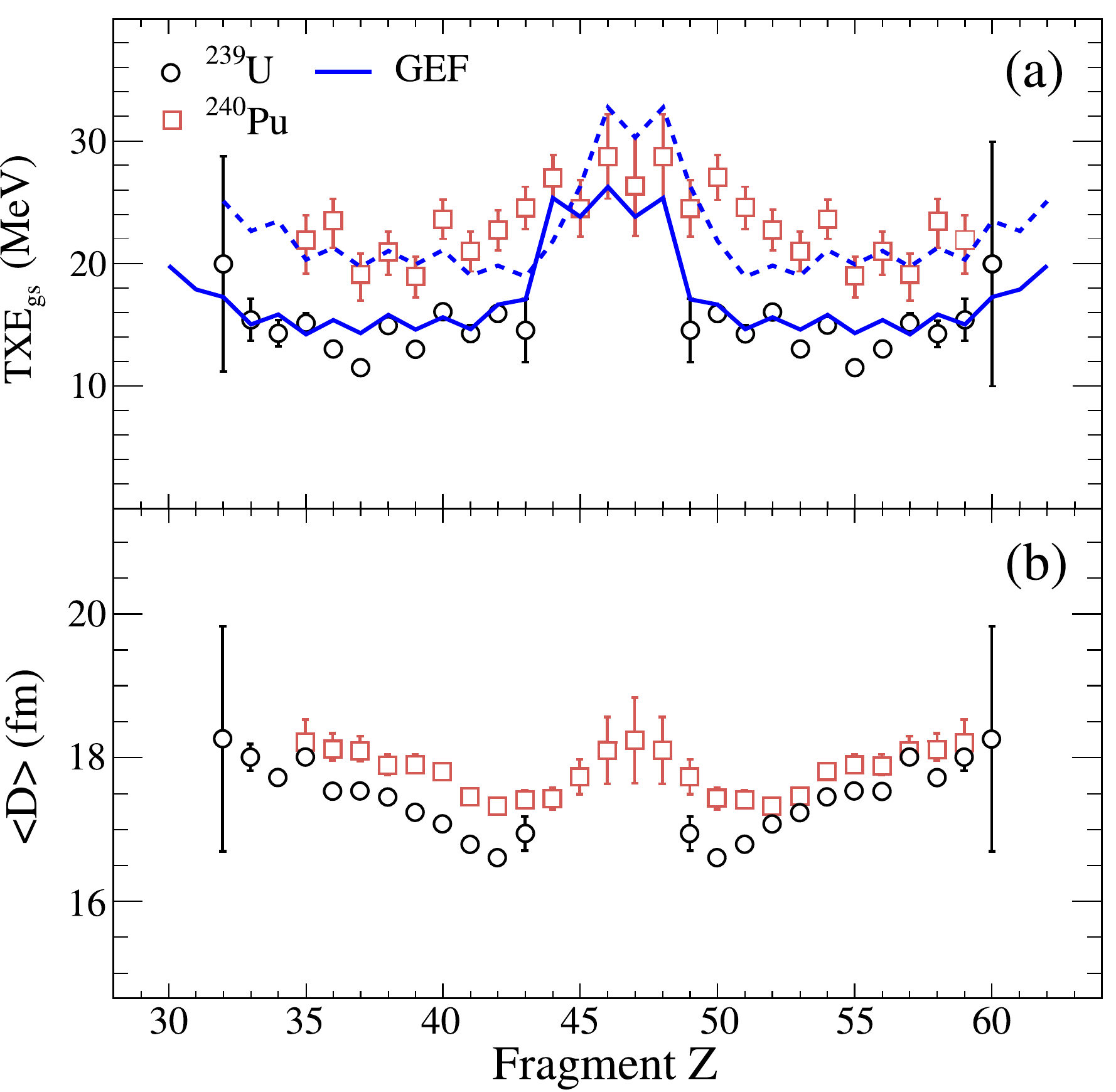}
\caption{Average total excitation energy and elongation at scission: (a) Total excitation energy of the fragments at scission with respect to the ground state of the fissioning system. Present data (black circles) are compared with a previous measurement of $^{240}$Pu~\cite{Caamano15} (red squares) and with GEF calculations (v. 2019/1.1)~\cite{GEF} (solid and dotted blue lines).  (b) Elongation of the fissioning system at scission as a function of the atomic number.}
\label{fig::TXE_COMP}
\end{figure}

Figure~\ref{fig::TXE_COMP} (a) presents the average total excitation energy of the fission fragments with respect to the ground state of the fissioning system ($\langle TXE_{g.s.}\rangle$) as a function of the atomic number of the fragments, determined as the difference between the $\langle TKE\rangle$ and the $\langle Q\rangle$-value of the reaction:
\begin{equation}
\langle TXE_{g.s.}\rangle(Z) = \langle Q\rangle(Z) - \langle TKE\rangle(Z).
\end{equation}
Both set of data, $^{239}$U (black circles) and $^{240}$Pu (red squares) are compared with GEF calculations (solid and dotted blue lines, respectively). Present data report systematically lower $\langle TXE_{g.s.}\rangle$ than that in $^{240}$Pu. Both data sets are in good agreement with the model in asymmetric splits, showing a rather constant value. At symmetry, the calculation predicts a sharp transition with a large $\langle TXE_{g.s.}\rangle$. This cannot be probed in the present data because of the lack of statistics at symmetry. Concerning $^{240}$Pu data, the enhancement of $\langle TXE_{g.s.}\rangle$ at the symmetry is observed with a smoother transition with respect to GEF. There is a clear even-odd oscillation produced by the $\langle Q\rangle$-value oscillation that is slightly underestimated by the model. 

This observation indicates that, for asymmetric fission, the potential energy of the system at the scission point strongly depends on the fissioning system but barely on the fission asymmetry. In $Z=50$, the lower excitation energy due to the sphericity of Sn, as indicated by the minimum neutron multiplicity, is compensated by a larger excitation energy in the light partner fragment.

Figure~\ref{fig::TXE_COMP} (b) shows the elongation of the fissioning system at scission ($D$), calculated from the Coulomb repulsion as:
\begin{equation}
\langle D\rangle = 1.44 \frac{Z_1Z_2}{\langle TKE\rangle}.
\end{equation}

Both systems, $^{239}$U (black circles) and $^{240}$Pu (red squares), show a minimum around $Z=50$. This observation is consistent with the low neutron multiplicity observed in Fig.~\ref{fig::NZ}, however, as discussed in Ref.~\cite{caamano17}, this minimum cannot be explained only by means of the low deformation of Sn but a shorter neck length between both fragments is needed. For this atomic-number split, the scission configuration of $^{239}$U is $\sim$1~fm more compact than $^{240}$Pu. This cannot be explained neither by the additional mass of $^{240}$Pu with respect to $^{239}$U that increases the nucleus size in $\sim$0.02~fm, using $d= 1.22[\textrm{fm}]\times(A_{light}^{1/3}+A_{heavy}^{1/3})$.

At larger asymmetry, both systems show rather similar elongations in the heavy fragment, $Z>51$, even presenting $\sim 5$ MeV of difference in $TXE$. Contrary, in the light region, the elongation is larger for $^{240}$Pu than for $^{239}$U. This observation suggests that, for asymmetric fission, the scission elongation is more sensitive to the heavy fragment than to $TXE$.

\section{CONCLUSIONS}
\label{sec_con}
In summary, the presented results are valuable constraints for current fission models. They provide new information such as the sharing of protons and neutrons at the scission point and the total kinetic and total excitation energy of the fragments by combining a large number of fission-fragment experimental observables.

The simultaneous measurement of the fission yields together with the velocity of the fission fragments of $^{239}$U confirms the impact of the intrinsic structure of Sn at scission: the neutron excess deviates from the liquid drop model at $Z\sim 50$ driven by $^{132}$Sn, although the effect on the yields is very reduced. The minimum neutron multiplicity around Sn suggests a low deformation that is consistent with the maximum total kinetic energy, indicating a compact configuration and a short neck.

The present results, together with previous data on $^{240}$Pu, can explain the fragment production with the combined effect of proton and neutron octupole shells: the relative yields around $Z=52,56$ shells depend on the simultaneous production of neutron shells $N=56,84,88$.

The $TXE$ shows that the excitation energy of fragments in asymmetric splits balanced out to render an almost flat behaviour. The different $TXE$ in U and Pu barely contributes to the scission elongation, which is mainly decided by the heavy fragment. 

The accuracy of the measurement is reflected in the good agreement with previous data and evaluations in both, fission yields and kinetic energies. The good agreement with the GEF code proves its strength in the calculation of fission-fragment observables and it restates this code as a useful tool for nuclear applications.   

\section*{ACKNOWLEDGMENTS}

The authors acknowledge the excellent support from the GANIL technical staff, J. Goupil, G. Fremont, L. Menager, J.A. Ropert, and C. Spitaels. This work was partially supported by the Spanish Ministry of Research and Innovation under the budget items FPA2015-71690-P and RYC-2012-11585, and by the Xunta de Galicia under item ED431F 2016/002. The research leading to these results has received funding from the European Union's HORIZON2020 Program under grant agreement n$^{\circ}$654002.

\Urlmuskip=0mu plus 1mu\relax
\bibliography{e753_paper2} 
\bibliographystyle{myapsrev4-1}
\end{document}